\documentclass[ALICE,manyauthors]{cernphprep}
\usepackage[comma,square,numbers,sort&compress]{natbib}

\usepackage{lineno}
\usepackage{xspace}
\usepackage{hyperref}
\usepackage{booktabs}
\usepackage[T1]{fontenc}
\usepackage{orcidlink}

\usepackage[capitalize]{cleveref}

\newcommand{\dndeta}       {\ensuremath{\mathrm{d}N_\mathrm{ch}/\mathrm{d}\eta}\xspace}
\newcommand{\rhokXi}{\ensuremath{\rho_{\Delta\Xi\Delta\mathrm{K}}}\xspace}

\begin{document}

\begin{titlepage}
\PHyear{2024}      
\PHnumber{151}     
\PHdate{27 May} 

\title{Probing strangeness hadronization with event-by-event production of multistrange hadrons}
\ShortTitle{Event-by-event production of multistrange hadrons}   

\Collaboration{ALICE Collaboration\thanks{See Appendix~\ref{app:collab} for the list of collaboration members}}
\ShortAuthor{ALICE Collaboration} 

\begin{abstract}
This Letter presents the first measurement of event-by-event fluctuations of the net number (difference between the particle and antiparticle multiplicities) of multistrange hadrons $\Xi^-$ and $\overline{\Xi}^+$ and its correlation with the net-kaon number using the data collected by the ALICE Collaboration in pp, p--Pb, and Pb--Pb collisions at a center-of-mass energy per nucleon pair $\sqrt{s_{\mathrm{NN}}}=5.02\ \mathrm{TeV}$.
The statistical hadronization model with a correlation over three units of rapidity between hadrons having the same and opposite strangeness content successfully describes the results.
On the other hand, string-fragmentation models that mainly correlate strange hadrons with opposite strange quark content over a small rapidity range fail to describe the data. 
\end{abstract}
\end{titlepage}

\setcounter{page}{2}


In high-energy hadronic and heavy-ion collisions, strange quarks are dominantly produced from gluon fusion~\cite{Rafelski:1982pu}. In Pb--Pb collisions at the Large Hadron Collider (LHC), a thermalized medium of deconfined partons, the strongly-interacting quark--gluon plasma (sQGP), is expected to form, where the efficient production of strange--antistrange quark pairs enables the thermal and chemical equilibration of strangeness in the medium~\cite{Rafelski:1982pu, Koch:1986ud}.
Various experimental results from high-multiplicity pp collisions at the LHC demonstrate striking similarities to results from Pb--Pb collisions. Notably, the ratios between strange and non-strange hadron yields show a smooth increase with increasing particle multiplicity across collision systems~\cite{ALICE:2016fzo, ALICE:2018pal, ALICE:2019avo}, and the patterns of multi-particle correlations in pp collisions closely resemble those seen in Pb--Pb collisions~\cite{ALICE:2012eyl, CMS:2016fnw, ALICE:2019zfl, ALICE:2023ulm}.
Theoretically, explaining such experimental observations, and the hadron production in general, requires phenomenological modeling of the hadronization process, as its inherently non-perturbative nature prevents us from performing reliable quantum chromodynamics (QCD) ab-initio calculations. Two different approaches, namely statistical hadronization~\cite{Hagedorn:1980kb, Hagedorn:1984uy} and Lund string fragmentation~\cite{Andersson:1983ia}, are commonly employed to address this problem. The statistical hadronization model (SHM) is based on a thermodynamic approach and the hadron abundances are determined at the freeze-out of inelastic interactions from the derivatives of the partition function of the system, which is assumed to be an ideal gas of hadrons and resonances (HRG) at local equilibrium. Event-by-event conservation laws are implemented using the Canonical Ensemble (CE) formulation of statistical mechanics~\cite{Bzdak:2012an, Braun-Munzinger:2020jbk}. Using the CE, the SHM model can describe the yield of light-flavored particles across all colliding systems with a precision better than 20\%~\cite{Vovchenko:2019kes, Cleymans:2020fsc}. The SHM implementations of Refs.~\cite{Vovchenko:2019kes, Cleymans:2020fsc} differ in the parametrizations of the system volume, $V$, and chemical freeze-out temperature, $T_{\mathrm{chem}}$, as well as in the accounting for a possible incomplete thermalization of the total strangeness at low multiplicity via a strangeness saturation factor, $\gamma_s$~\cite{Letessier:1993hi}. On the other hand, in the string-fragmentation picture, as implemented in PYTHIA~\cite{Bierlich:2022pfr}, final-state partons are connected by color flux tubes, known as strings, which break up into smaller segments for large string lengths. Additional quark--antiquark pairs are produced during this process: once no more energy is available for further splitting, hadrons are formed.
The starting string configuration for hadronization is determined by the arrangement of opposite colors and anticolors according to color reconnection mechanisms~\cite{Skands:2014pea, Christiansen:2015yqa, Bierlich:2022pfr}.
Adding further interactions between the strings leading to the formation of baryon junctions~\cite{Christiansen:2015yqa} and ropes~\cite{Bierlich:2014xba}, PYTHIA can describe the multiplicity dependence of the ratio between the yields of strange and non-strange particles~\cite{Nayak:2018xip}.
Consequently, new observables are needed to discriminate between these two approaches.

String fragmentation and canonical statistical hadronization provide different treatments of the conservation laws. In the former, quantum numbers are conserved at a local level because of the formation of quark--antiquark pairs in the string breaking process, while in the latter, conservation laws hold over a finite correlation volume, $V_{\mathrm{c}}$~\cite{Vovchenko:2019kes}. Consequently, canonical charge conservation implies correlations between any two hadrons carrying either same- or opposite-sign quantum numbers, showing a decreasing correlation strength for increasing correlation volume.
Instead, in the string fragmentation model, a strong correlation exists mostly between oppositely-charged hadrons because of the quantum number conservation at each string breaking. In principle, such an effect has no significant multiplicity dependence, except from that coming from specific implementations of the color reconnection mechanism.

The difference in the quantum number conservation between these two models can be probed by analyzing the event-by-event correlation between different hadron species. This approach has previously been applied by the ALICE Collaboration to study baryon number conservation via net-proton fluctuations~\cite{ALICE:2019nbs}. There, it was concluded that baryon number is conserved through long rapidity-range correlations. In this Letter, the first study in this regard in the strangeness sector is presented analyzing the event-by-event correlation between charged kaons and $\Xi^-$ and $\overline{\Xi}^+$ baryons. In the following, $\Xi$ is used instead of $\Xi^-$ and $\overline{\Xi}^+$ for brevity, unless otherwise specified. These species are chosen because they are minimally affected by correlations other than those induced by the quantum number conservation, e.g. by the decay of heavier states into charged kaons and $\Xi$ baryons. In addition, their production is only marginally affected by weak feed-down, which comes from $\Omega$-baryon decays. The observables considered in this Letter are the normalized second-order cumulant of net-$\Xi$ number and the correlation between net-$\Xi$ and net-kaon numbers. The net-particle numbers are defined in terms of the event-by-event multiplicities, $n$, of particles and antiparticles as $n_{\Delta\mathrm{K}}=n_{\mathrm{K}^+}-n_{\mathrm{K}^-}$ and $n_{\Delta\Xi}=n_{\overline{\Xi}^+}-n_{\Xi^-}$ for charged kaons and charged $\Xi$ baryons, respectively. The normalized second-order cumulant of net-$\Xi$ number and the correlation between net-$\Xi$ and net-kaon numbers are defined as

\begin{linenomath}
\begin{equation}\label{eq:cumulant_ratio}
\dfrac{\kappa_2(\Delta\Xi)}{\kappa_1(\overline{\Xi}^+ + \Xi^-)}=\dfrac{\kappa_2(\overline{\Xi}^+) + \kappa_2(\Xi^-)-2\kappa_{11}(\overline{\Xi}^+,\Xi^-)}{\kappa_1(\overline{\Xi}^+ + \Xi^-)},
\end{equation}
\begin{equation}\label{eq:rho}
\rho_{\Delta\Xi\Delta\mathrm{K}}=\dfrac{\kappa_{11}(\overline{\Xi}^+, K^+)+\kappa_{11}(\Xi^-, K^-)-\kappa_{11}(\overline{\Xi}^+, K^-)-\kappa_{11}(\Xi^-, K^+)}{\sqrt{\kappa_2(\Delta\Xi)\kappa_2(\Delta K)}},
\end{equation}
\end{linenomath}
respectively, where
\begin{equation}
    \kappa_1(\mathrm{A})=\langle n_\mathrm{A}\rangle,
\end{equation}
\begin{equation}
    \kappa_2(\mathrm{A})=\langle n_\mathrm{A}^2\rangle - \langle n_\mathrm{A} \rangle^2,
\end{equation}
\begin{equation}
    \kappa_{11}(\mathrm{A},\mathrm{B})=\langle n_\mathrm{A}n_\mathrm{B}\rangle-\langle n_\mathrm{A}\rangle\langle n_\mathrm{B}\rangle,
\end{equation}

and $n_{\mathrm{A,B}}$ and $\langle n_\mathrm{A,B} \rangle$ indicate the event-by-event and event-averaged number of particles of species $\mathrm{A}$ or $\mathrm{B}$, respectively. The $m$-th order cumulants of net particles are denoted as $\kappa_m$ in \cref{eq:cumulant_ratio} and \cref{eq:rho}. The normalized second-order cumulant is only affected by opposite strangeness sign correlation, whereas both the correlations of same- and opposite-strangeness pairs have an impact on $\rho_{\Delta\Xi\Delta\mathrm{K}}$. Both observables are sensitive to the locality of strangeness conservation, which affects the magnitude of the correlations. Therefore, these observables are powerful tools to distinguish among different hadronization scenarios. In addition, the studied quantities are independent of volume fluctuations under the hypothesis of particle--antiparticle balance at the energies available at the LHC from small to large systems~\cite{Braun-Munzinger:2016yjz, ALICE:2023ulv}.

The results reported in this Letter are extracted from data collected by the ALICE Collaboration in pp, p--Pb, and Pb--Pb collisions at a center-of-mass energy per nucleon--nucleon pair $\sqrt{s_{\mathrm{NN}}}=5.02\ \mathrm{TeV}$. The ALICE apparatus and its performance are described in detail in Refs.~\cite{ALICE:2008ngc, ALICE:2014sbx}.
Events are selected with a minimum bias (MB) trigger based on a coincidence of signals in the two V0 scintillator arrays~\cite{ALICE:2013axi}, which are placed at both sides of the nominal interaction point, covering the pseudorapidity, $\eta$, intervals $-3.7<\eta<-1.7$ and $2.8<\eta<5.1$.
In the Pb--Pb sample, an additional trigger based on the amplitude of the V0 signal is applied to enhance the selection of central (head-on) and semi-central collisions. Further selections are applied to reject pile-up events~\cite{Arslandok:2022dyb}. Finally, the position of the reconstructed primary vertex (PV) along the beam direction ($z$-axis) is required to be within 10 cm around the nominal interaction point. After event selections, the datasets consist of about $900\times10^6$ pp collisions, $600\times10^6$ p--Pb collisions, and $400\times10^6$ Pb--Pb collisions. The selected samples are subdivided into multiplicity classes defined according to the signal amplitudes measured in the V0 detectors~\cite{ALICE:2018tvk}.

Charged particles are tracked in the central-barrel detectors of ALICE, which cover the full azimuth in the midrapidity region. The pseudorapidity acceptance of the present measurement is $\vert\eta\vert<0.8$. The charged kaon and charged $\Xi$ candidates are selected in the transverse momentum ($p_{\mathrm{T}}$) ranges $0.2<p_{\mathrm{T}}<1.0\ \mathrm{GeV}/c$ and $1.0<p_{\mathrm{T}}<3.0\ \mathrm{GeV}/c$, respectively, to cover the bulk of the production. The purity of the selected samples of candidates in these momentum intervals is $\gtrsim 95\%$. Charged kaons are directly tracked in the detectors, while $\Xi$ baryons are reconstructed via their weak decay to a charged pion and a $\Lambda$ baryon: the latter is identified via its weak decay into a proton and a charged pion, $\Xi^-\to \pi^- + \Lambda(\to \mathrm{p} + \pi^-)$. The charge-conjugate decay is employed for $\overline{\Xi}^+$.

The reconstruction of tracks is based on the space points measured in the Inner Tracking System (ITS)~\cite{ALICE:2010tia} and the Time Projection Chamber (TPC)~\cite{Alme:2010ke}. Track selections are applied to ensure a good quality of the track reconstruction~\cite{ALICE:2013wgn, ALICE:2019hno}.
To avoid intersections in the samples of tracks used for the reconstruction of kaons and $\Xi$, which might produce spurious correlations, complementary selections are applied to the distance of closest approach (DCA) of tracks to the PV. Specifically, $\vert\mathrm{DCA}\vert<0.1\ \mathrm{cm}$ for kaons, while $\vert\mathrm{DCA}\vert>0.1\ \mathrm{cm}$ for $\Xi$-decay products. Tracked particles are identified by measuring their specific energy loss, $\mathrm{d}E/\mathrm{d}x$, in the TPC gas volume. For all candidates, the difference between the measured TPC $\mathrm{d}E/\mathrm{d}x$ signal and the one expected for the considered particle species is required to be less than $3\sigma$, where $\sigma$ is the resolution on the $\mathrm{d}E/\mathrm{d}x$ in the data assuming a Gaussian shape for the TPC particle identification (PID) signal. For kaons, additional particle selection criteria are applied: for candidates with $p_{\mathrm{T}}<0.4\ \mathrm{GeV}/c$ in Pb--Pb collisions, the $\mathrm{d}E/\mathrm{d}x$ measured with the ITS is used to improve the rejection of electrons and positrons; for $p_{\mathrm{T}}>0.4\ \mathrm{GeV}/c$, the particle velocity measured with the Time-Of-Flight (TOF) detector is employed to reject charged pions and muons: the discrepancy between the measured velocity and the one expected for kaons is requested to be smaller than~$3\sigma$.
The algorithm used to reconstruct the $\Xi$-decay vertices from tracks is similar to that used in previous studies~\cite{ALICE:2012yqk, ALICE:2013xmt, ALICE:2015mpp}. A large fraction of combinatorial background contaminates the sample of selected $\Xi$ candidates. To address this, a machine learning (ML) method is used to enhance the signal-over-background ratio of the sample, as detailed in Ref.~\cite{ALICE-PUBLIC-2023-003}. The training of the ML algorithm is based on the cascade and two-body decay topological variables of the $\Xi$ and $\Lambda$ candidates, respectively.
A few examples of the invariant-mass distribution of the selected $\Xi$ and of the TPC and TOF PID variables of the selected kaons are reported in Ref.~\cite{ALICE-PUBLIC-2023-003}. In the 10\% most central Pb--Pb collisions, the average number of selected $\overline{\Xi}^++\Xi^-$ and $\mathrm{K}^++\mathrm{K^-}$ is about 0.07 and 16 per collision, respectively, before applying efficiency corrections.

The observables defined in \cref{eq:cumulant_ratio} and \cref{eq:rho} are corrected for the candidate-selection efficiency assuming a detector response with binomial fluctuations. The analytic expression for the efficiency correction of first- and second-order cumulants were obtained in previous works~\cite{Nonaka:2017kko}. The efficiencies are calculated using MC simulations in which particles produced by an event generator (PYTHIA 8 with Monash tune~\cite{Skands:2014pea, Bierlich:2022pfr} for pp, EPOS LHC~\cite{Pierog:2013ria} for p--Pb, and HIJING~\cite{Wang:1991hta} for Pb--Pb) are transported through an accurate model of the ALICE apparatus via GEANT4~\cite{GEANT4:2002zbu}. The efficiencies are $p_{\mathrm{T}}$ and multiplicity dependent, ranging from 1\% to 8\% for $\Xi$ baryons and from 5\% to 30\% for kaons. The resulting correction factors on $\rho_{\Delta\Xi\Delta\mathrm{K}}$ and $\kappa_2(\Delta\Xi)/\kappa_1(\overline{\Xi}^+ + \Xi^-)$ are 10 to 15 and 0.95, respectively. Using a closure test based on MC simulations, it was checked that the efficiency-correction procedure does not introduce any significant bias in the corrected cumulants.

The statistical uncertainties are estimated using the subsample method~\cite{Pandav:2018bdx}. The systematic uncertainties are obtained by extracting cumulants and correlations using different variations of the candidate selection criteria.
This procedure is repeated using several combinations of the different variations~\cite{ALICE-PUBLIC-2023-003}. The systematic uncertainty associated to each of the sources are reported in \cref{app:sys}. The total systematic uncertainty in the cumulants is computed, for each multiplicity class, as the standard deviation of the results obtained with the different combinations. The average of the multi-trial results is employed as the central value of the observable. The systematic uncertainties are fully correlated across different multiplicity intervals.

\begin{figure*}[t]
    \centering
    \includegraphics[width=0.49\textwidth]{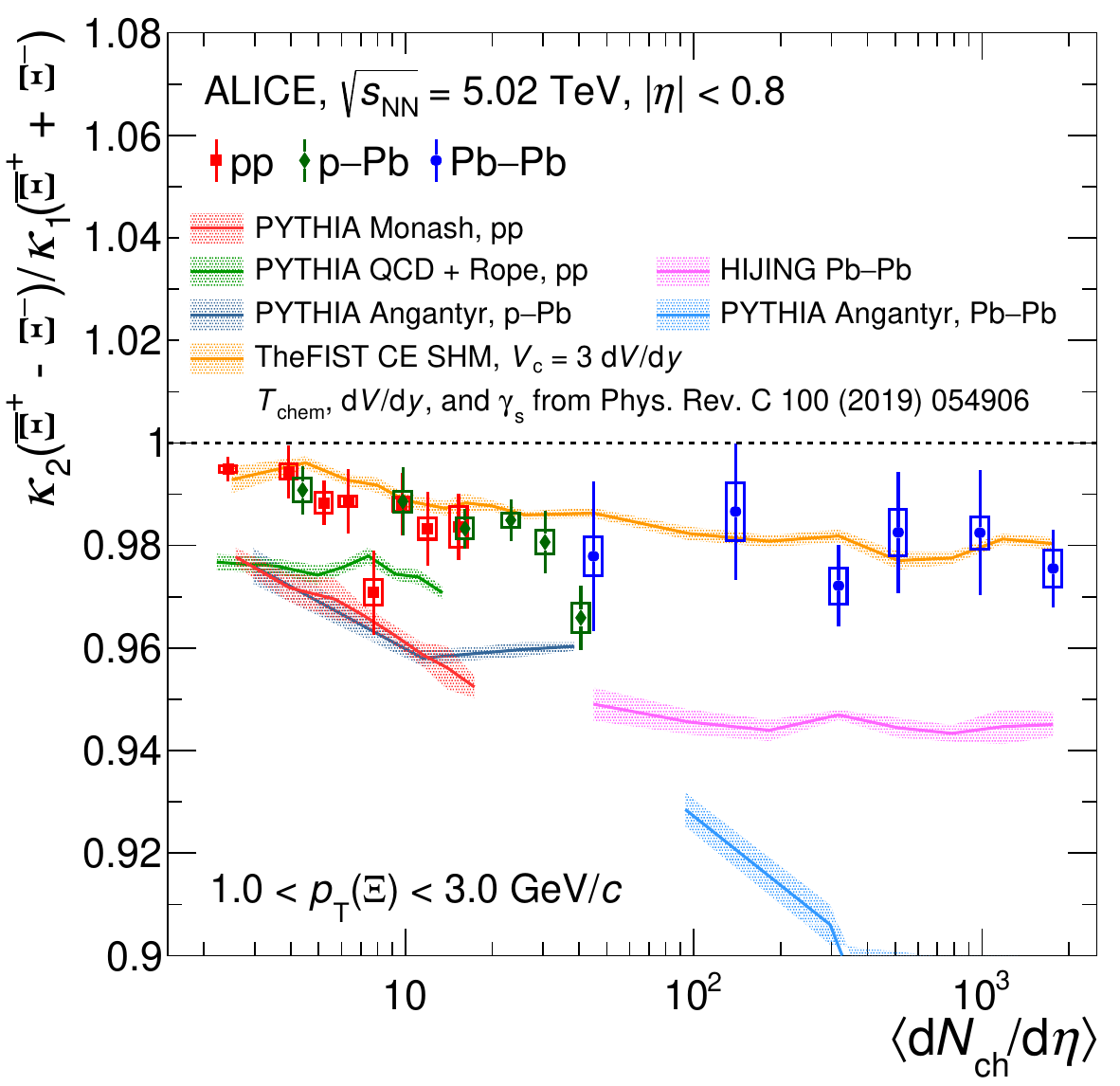}
    \includegraphics[width=0.49\textwidth]{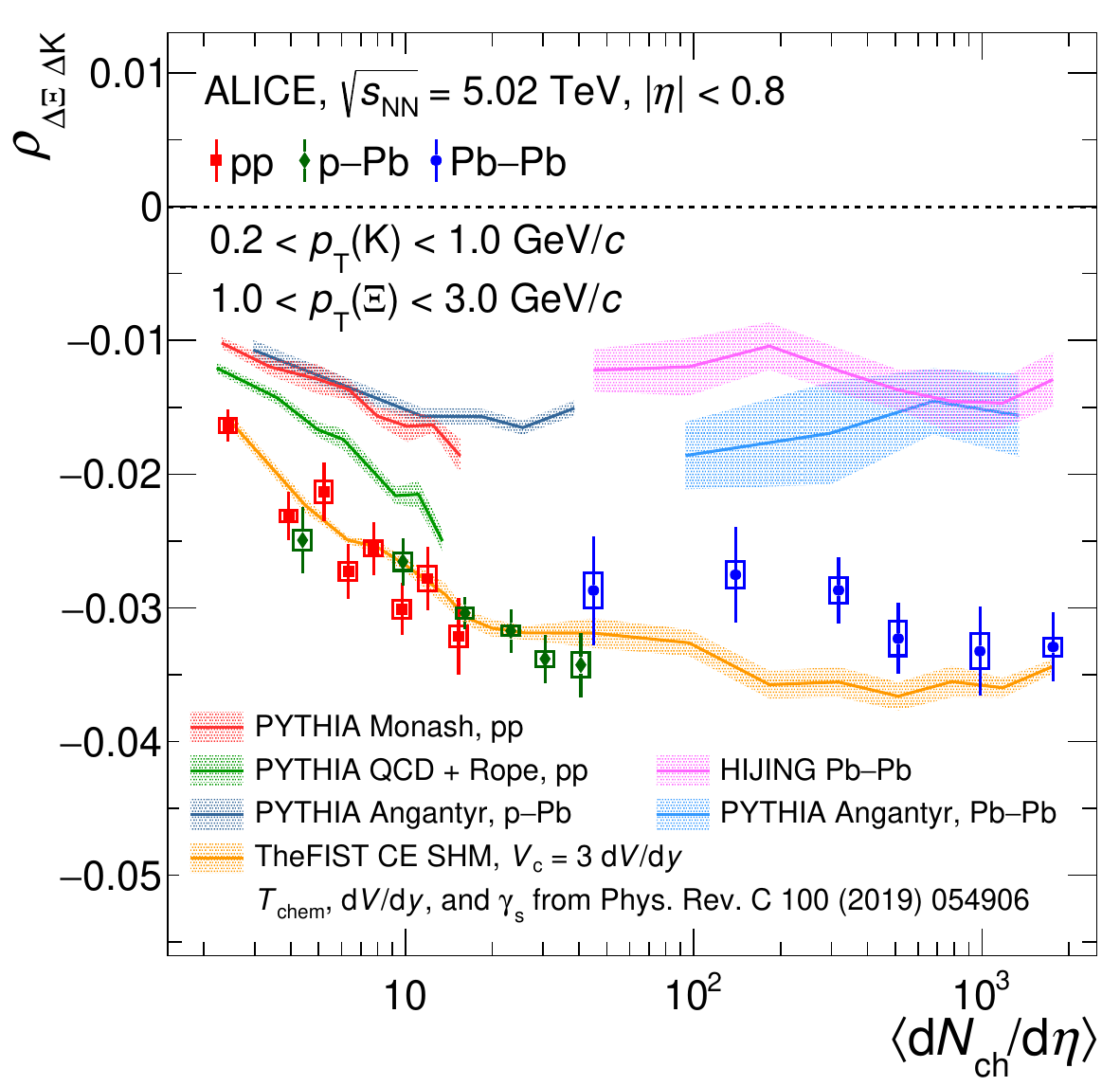}

    \caption{
    Normalized second order cumulant of net $\Xi$ (left panel) and correlation between net $\Xi$ and net kaon (right panel), as a function of the average multiplicity at midrapidity, in pp (red squares), p--Pb (green diamonds), and Pb--Pb collisions (blue circles) at $\sqrt{s_{\mathrm{NN}}}=5.02\ \mathrm{TeV}$. Statistical and systematic uncertainties are shown via error bars and boxes, respectively. The experimental measurements are compared to several model predictions, shown as bands. The width of the bands represents the statistical uncertainty of the predictions. The average multiplicity values are obtained from~Refs.~\cite{ALICE:2013wgn, ALICE:2015juo, ALICE:2021ovi}: their uncertainties are not shown in the plot.}
    \label{fig:final_plot}
\end{figure*}

The normalized second-order cumulant of net $\Xi$ and the correlation between net-kaon and net-$\Xi$ numbers are shown in the left and right panels of~\cref{{fig:final_plot}}, respectively. The observables are shown as a function of the average charged-particle multiplicity at midrapidity, $\langle \mathrm{d}N_{\mathrm{ch}} / \mathrm{d}\eta \rangle$, enabling the comparison of the results from different colliding systems. The results show a continuous evolution as a function of the multiplicity from pp to Pb--Pb collisions for the correlation term and the normalized second-order cumulant. The measurements are compared to the SHM and Lund string-fragmentation model, to probe the correlation volume between strange hadrons and the presence of same-sign correlations originating from the different treatments of net-strangeness conservation. The width of the bands in~\cref{{fig:final_plot}} depicts the statistical uncertainty of the MC simulations corresponding to the different models. A long-range rapidity correlation implies a smaller deviation from the Poisson baseline,
corresponding to the Grand Canonical ensemble (GCE) limit, than a short-range correlation, which is generally present in string models. For the normalized second-order cumulant of net $\Xi$, the Poisson baseline equals unity, while it is zero for $\rho_{\Delta\Xi\Delta\mathrm{K}}$. The predictions of the SHM within the CE framework, shown in~\cref{{fig:final_plot}},
are obtained with the Thermal-FIST package~\cite{Vovchenko:2019pjl}. The model parameters, such as the chemical freeze-out temperature, $T_{\mathrm{chem}}$, and the volume per unit of rapidity, $\mathrm{d}V/\mathrm{d}y$, are tuned using the hadron yields measured by ALICE and setting the correlation volume to $V_{\mathrm{c}}=3\ \mathrm{d}V/\mathrm{d}y$. The model includes a strangeness saturation parameter, $\gamma_s$, which accounts for incomplete total strangeness equilibration at $\langle\dndeta\rangle\lesssim100$~\cite{Vovchenko:2019kes}. This parameter is needed in Thermal-FIST to describe the average yields of (multi)strange hadrons but it does not have any significant effect on the observables shown in~\cref{fig:final_plot}. This parametrization allows us to describe both the normalized second-order cumulant of net $\Xi$ and $\rho_{\Delta\Xi\Delta\mathrm{K}}$ within uncertainties, with a slight tension in semicentral Pb--Pb collisions for $\rho_{\Delta\Xi\Delta\mathrm{K}}$.
Predictions from the Lund string-fragmentation models for the high-multiplicity regime corresponding to heavy-ion collisions are obtained with HIJING~\cite{Wang:1991hta} and PYTHIA Angantyr~\cite{Bierlich:2018xfw}. The Angantyr model is also used for p--Pb collisions,
while for pp collisions, the PYTHIA model with different color reconnection (CR) schemes is used~\cite{Bierlich:2022pfr}. The predictions shown in~\cref{{fig:final_plot}} are obtained either with the multiparton-interaction (MPI) based CR (Monash tune~\cite{Skands:2014pea}) or with a QCD-based CR approach~\cite{Christiansen:2015yqa}. For the latter calculation, the effect of rope hadronization, in which spatially overlapping strings are allowed to combine into ropes with a larger effective string tension, is also included.
Rope hadronization is responsible for an enhancement in the production of strangeness~\cite{Christiansen:2015yqa}, while the QCD CR describes an increase in baryon production due to the formation of baryon junctions~\cite{Bierlich:2014xba}.

The normalized second-order cumulants of net $\Xi$ are consistently below unity over the entire multiplicity range, which can be understood as an effect of quantum number (baryon, strangeness) conservation~\cite{Vovchenko:2020kwg}. The results are consistent with the SHM with a correlation range of three units of rapidity throughout the analyzed collision systems. In the context of the SHM, this suggests the presence of a long-range rapidity correlation between two hadrons with opposite strangeness contents due to the conservation of strangeness. On the other hand the different kinds of string models overestimate the strength of the correlation, which is quantified by the deviation of the predictions from unity.

Furthermore, the measured correlation between net kaon and net $\Xi$, \rhokXi, is sensitive to the range of the correlation due to strangeness conservation and to the possible correlation between hadrons with same-sign strange quantum numbers. A significant anti-correlation between net kaon and net $\Xi$ is observed across all collision systems.
Even though the string fragmentation model has small-range rapidity correlation, the SHM predicts a more significant deviation from the Poisson baseline with respect to the string fragmentation model, as the latter does not include any significant correlations among hadrons with same-sign strangeness. The SHM prediction describes the measured \rhokXi, indicating a sizeable contribution of same strangeness sign correlation.

From the simultaneous fit of the \rhokXi and normalized second-order cumulant for net $\Xi$ in Pb--Pb collisions using the Thermal-FIST package, the correlation volume is determined to be $V_{\mathrm{c}}=3.19\pm0.14\ \mathrm{d}V/\mathrm{d}y$, with a fit probability of $P=0.94$ (see Appendix~\ref{app:vc_determination} for further details). This volume, valid only for Pb--Pb collisions, is compatible within $1.4\sigma$ with a value $V_{\mathrm{c}}=3\ \mathrm{d}V/\mathrm{d}y$, which was obtained in previous analyses of hadron yield ratios within the canonical statistical model across different colliding systems~\cite{Vovchenko:2019kes}. In the context of the SHM, this result shows that a large correlation volume regulates the conservation of strangeness, implying that correlations are formed at earlier times than predicted by string fragmentation. A large correlation volume for baryon number conservation was also observed from the study of net-proton fluctuations. See a comparison between the experimental data from the ALICE Collaboration and the predictions obtained with Thermal-FIST in Appendix~\ref{app:netproton}. On the other hand, a correlation volume of $1.6\ \mathrm{d}V/\mathrm{d}y$ was obtained from the analysis of event-by-event fluctuations of antideuterons~\cite{ALICE:2022xiu}, possibly because the formation of bound objects such as light nuclei differs from the production of other light-flavor hadrons. Specifically, these results might indicate later formation times of light nuclei compared to other hadron species~\cite{Sun:2022rjh}.

In summary, the simultaneous measurement of both the presented observables has a high discriminative power against the different model predictions. The correlation and normalized second-order cumulant measurements are well described by the CE SHM formalism. The model reproduces both the absolute value of the observables and their multiplicity trends. These results are consistent with previous work~\cite{Vovchenko:2019kes}, in which the yield ratios of charged particles, such as $\Xi/\pi$, in pp, p--Pb, and Pb--Pb collisions were also studied within the framework of the canonical statistical model implemented in Thermal-FIST. All of the tested predictions based on non-thermally equilibrated systems, and string fragmentation, qualitatively describe a negative correlation of net-$\Xi$ and net-kaon numbers, along with a normalized second-order cumulant smaller than one. However, they fail to quantitatively describe the experimental results, both in the low- and high-multiplicity regions. The PYTHIA 8 tune closest to the experimental results is the QCD-based CR approach with rope hadronization: this model can describe the hadron yield ratios at low multiplicity, as shown in previous studies ~\cite{Nayak:2018xip}. Nevertheless, a significant combined deviation of $7.5\sigma$ between the data in pp collisions and the string fragmentation predictions is observed. Such a discrepancy could be resolved only by simultaneously including long-range and same-sign correlations in the string fragmentation framework. Such long-range correlations are a feature of the thermally-equilibrated system modeled by Thermal-FIST that successfully describe the measurements. It is possible that for cumulants higher than the second order, deviations from the thermal baseline might occur, as these are associated with different relaxation times to reach thermalization~\cite{Gupta:2022phu} or due to the presence of a phase transition~\cite{Bazavov:2017dus}.
Such a scenario could be probed in the future, with the ongoing LHC Run 3 data, extending such studies to higher order cumulants.
Furthermore, the measurements provided here can also be used to probe the correlation volume in more elaborated statistical hadronization models such as in Ref.~\cite{Braun-Munzinger:2023gsd}.


\newenvironment{acknowledgement}{\relax}{\relax}
\begin{acknowledgement}
\section*{Acknowledgements}

The ALICE Collaboration would like to thank all its engineers and technicians for their invaluable contributions to the construction of the experiment and the CERN accelerator teams for the outstanding performance of the LHC complex.
The ALICE Collaboration gratefully acknowledges the resources and support provided by all Grid centres and the Worldwide LHC Computing Grid (WLCG) collaboration.
The ALICE Collaboration acknowledges the following funding agencies for their support in building and running the ALICE detector:
A. I. Alikhanyan National Science Laboratory (Yerevan Physics Institute) Foundation (ANSL), State Committee of Science and World Federation of Scientists (WFS), Armenia;
Austrian Academy of Sciences, Austrian Science Fund (FWF): [M 2467-N36] and Nationalstiftung f\"{u}r Forschung, Technologie und Entwicklung, Austria;
Ministry of Communications and High Technologies, National Nuclear Research Center, Azerbaijan;
Conselho Nacional de Desenvolvimento Cient\'{\i}fico e Tecnol\'{o}gico (CNPq), Financiadora de Estudos e Projetos (Finep), Funda\c{c}\~{a}o de Amparo \`{a} Pesquisa do Estado de S\~{a}o Paulo (FAPESP) and Universidade Federal do Rio Grande do Sul (UFRGS), Brazil;
Bulgarian Ministry of Education and Science, within the National Roadmap for Research Infrastructures 2020-2027 (object CERN), Bulgaria;
Ministry of Education of China (MOEC) , Ministry of Science \& Technology of China (MSTC) and National Natural Science Foundation of China (NSFC), China;
Ministry of Science and Education and Croatian Science Foundation, Croatia;
Centro de Aplicaciones Tecnol\'{o}gicas y Desarrollo Nuclear (CEADEN), Cubaenerg\'{\i}a, Cuba;
Ministry of Education, Youth and Sports of the Czech Republic, Czech Republic;
The Danish Council for Independent Research | Natural Sciences, the VILLUM FONDEN and Danish National Research Foundation (DNRF), Denmark;
Helsinki Institute of Physics (HIP), Finland;
Commissariat \`{a} l'Energie Atomique (CEA) and Institut National de Physique Nucl\'{e}aire et de Physique des Particules (IN2P3) and Centre National de la Recherche Scientifique (CNRS), France;
Bundesministerium f\"{u}r Bildung und Forschung (BMBF) and GSI Helmholtzzentrum f\"{u}r Schwerionenforschung GmbH, Germany;
General Secretariat for Research and Technology, Ministry of Education, Research and Religions, Greece;
National Research, Development and Innovation Office, Hungary;
Department of Atomic Energy Government of India (DAE), Department of Science and Technology, Government of India (DST), University Grants Commission, Government of India (UGC) and Council of Scientific and Industrial Research (CSIR), India;
National Research and Innovation Agency - BRIN, Indonesia;
Istituto Nazionale di Fisica Nucleare (INFN), Italy;
Japanese Ministry of Education, Culture, Sports, Science and Technology (MEXT) and Japan Society for the Promotion of Science (JSPS) KAKENHI, Japan;
Consejo Nacional de Ciencia (CONACYT) y Tecnolog\'{i}a, through Fondo de Cooperaci\'{o}n Internacional en Ciencia y Tecnolog\'{i}a (FONCICYT) and Direcci\'{o}n General de Asuntos del Personal Academico (DGAPA), Mexico;
Nederlandse Organisatie voor Wetenschappelijk Onderzoek (NWO), Netherlands;
The Research Council of Norway, Norway;
Pontificia Universidad Cat\'{o}lica del Per\'{u}, Peru;
Ministry of Science and Higher Education, National Science Centre and WUT ID-UB, Poland;
Korea Institute of Science and Technology Information and National Research Foundation of Korea (NRF), Republic of Korea;
Ministry of Education and Scientific Research, Institute of Atomic Physics, Ministry of Research and Innovation and Institute of Atomic Physics and Universitatea Nationala de Stiinta si Tehnologie Politehnica Bucuresti, Romania;
Ministry of Education, Science, Research and Sport of the Slovak Republic, Slovakia;
National Research Foundation of South Africa, South Africa;
Swedish Research Council (VR) and Knut \& Alice Wallenberg Foundation (KAW), Sweden;
European Organization for Nuclear Research, Switzerland;
Suranaree University of Technology (SUT), National Science and Technology Development Agency (NSTDA) and National Science, Research and Innovation Fund (NSRF via PMU-B B05F650021), Thailand;
Turkish Energy, Nuclear and Mineral Research Agency (TENMAK), Turkey;
National Academy of  Sciences of Ukraine, Ukraine;
Science and Technology Facilities Council (STFC), United Kingdom;
National Science Foundation of the United States of America (NSF) and United States Department of Energy, Office of Nuclear Physics (DOE NP), United States of America.
In addition, individual groups or members have received support from:
Czech Science Foundation (grant no. 23-07499S), Czech Republic;
European Research Council (grant no. 950692), European Union;
ICSC - Centro Nazionale di Ricerca in High Performance Computing, Big Data and Quantum Computing, European Union - NextGenerationEU;
Academy of Finland (Center of Excellence in Quark Matter) (grant nos. 346327, 346328), Finland.    
\end{acknowledgement}

\bibliographystyle{utphys} 

\bibliography{bibdb}

\providecommand{\href}[2]{#2}\begingroup\raggedright\begin{thebibliography}{10}

\bibitem{Rafelski:1982pu}
J.~Rafelski and B.~Muller, ``{Strangeness Production in the Quark - Gluon
  Plasma}'', \href{https://doi.org/10.1103/PhysRevLett.48.1066}{{\em Phys. Rev.
  Lett.} {\bfseries 48} (1982) 1066}. [Erratum: Phys.Rev.Lett. 56, 2334
  (1986)].

\bibitem{Koch:1986ud}
P.~Koch, B.~Muller, and J.~Rafelski, ``{Strangeness in Relativistic Heavy Ion
  Collisions}'', \href{https://doi.org/10.1016/0370-1573(86)90096-7}{{\em Phys.
  Rept.} {\bfseries 142} (1986) 167--262}.

\bibitem{ALICE:2016fzo}
{\bfseries ALICE} Collaboration, J.~Adam {\em et~al.}, ``{Enhanced production
  of multi-strange hadrons in high-multiplicity proton-proton collisions}'',
  \href{https://doi.org/10.1038/nphys4111}{{\em Nature Phys.} {\bfseries 13}
  (2017) 535--539}, \href{https://arxiv.org/abs/1606.07424}{{\ttfamily
  arXiv:1606.07424 [nucl-ex]}}.

\bibitem{ALICE:2018pal}
{\bfseries ALICE} Collaboration, S.~Acharya {\em et~al.}, ``{Multiplicity
  dependence of light-flavor hadron production in pp collisions at $\sqrt{s}$ =
  7 TeV}'', \href{https://doi.org/10.1103/PhysRevC.99.024906}{{\em Phys. Rev.
  C} {\bfseries 99} (2019) 024906},
  \href{https://arxiv.org/abs/1807.11321}{{\ttfamily arXiv:1807.11321
  [nucl-ex]}}.

\bibitem{ALICE:2019avo}
{\bfseries ALICE} Collaboration, S.~Acharya {\em et~al.}, ``{Multiplicity
  dependence of (multi-)strange hadron production in proton-proton collisions
  at $\sqrt{s}$ = 13 TeV}'',
  \href{https://doi.org/10.1140/epjc/s10052-020-7673-8}{{\em Eur. Phys. J. C}
  {\bfseries 80} (2020) 167},
  \href{https://arxiv.org/abs/1908.01861}{{\ttfamily arXiv:1908.01861
  [nucl-ex]}}.

\bibitem{ALICE:2012eyl}
{\bfseries ALICE} Collaboration, B.~Abelev {\em et~al.}, ``{Long-range angular
  correlations on the near and away side in p--Pb collisions at
  $\sqrt{s_{\mathrm{NN}}}=5.02$ TeV}'',
  \href{https://doi.org/10.1016/j.physletb.2013.01.012}{{\em Phys. Lett. B}
  {\bfseries 719} (2013) 29--41},
  \href{https://arxiv.org/abs/1212.2001}{{\ttfamily arXiv:1212.2001
  [nucl-ex]}}.

\bibitem{CMS:2016fnw}
{\bfseries CMS} Collaboration, V.~Khachatryan {\em et~al.}, ``{Evidence for
  collectivity in pp collisions at the LHC}'',
  \href{https://doi.org/10.1016/j.physletb.2016.12.009}{{\em Phys. Lett. B}
  {\bfseries 765} (2017) 193--220},
  \href{https://arxiv.org/abs/1606.06198}{{\ttfamily arXiv:1606.06198
  [nucl-ex]}}.

\bibitem{ALICE:2019zfl}
{\bfseries ALICE} Collaboration, S.~Acharya {\em et~al.}, ``{Investigations of
  Anisotropic Flow Using Multiparticle Azimuthal Correlations in pp, p-Pb,
  Xe-Xe, and Pb-Pb Collisions at the LHC}'',
  \href{https://doi.org/10.1103/PhysRevLett.123.142301}{{\em Phys. Rev. Lett.}
  {\bfseries 123} (2019) 142301},
  \href{https://arxiv.org/abs/1903.01790}{{\ttfamily arXiv:1903.01790
  [nucl-ex]}}.

\bibitem{ALICE:2023ulm}
{\bfseries ALICE} Collaboration, S.~Acharya {\em et~al.}, ``{Emergence of
  Long-Range Angular Correlations in Low-Multiplicity Proton-Proton
  Collisions}'', \href{https://doi.org/10.1103/PhysRevLett.132.172302}{{\em
  Phys. Rev. Lett.} {\bfseries 132} (2024) 172302},
  \href{https://arxiv.org/abs/2311.14357}{{\ttfamily arXiv:2311.14357
  [nucl-ex]}}.

\bibitem{Hagedorn:1980kb}
R.~Hagedorn and J.~Rafelski, ``{Hot Hadronic Matter and Nuclear Collisions}'',
  \href{https://doi.org/10.1016/0370-2693(80)90566-3}{{\em Phys. Lett. B}
  {\bfseries 97} (1980) 136}.

\bibitem{Hagedorn:1984uy}
R.~Hagedorn and K.~Redlich, ``{Statistical Thermodynamics in Relativistic
  Particle and Ion Physics: Canonical or Grand Canonical?}'',
  \href{https://doi.org/10.1007/BF01436508}{{\em Z. Phys. C} {\bfseries 27}
  (1985) 541}.

\bibitem{Andersson:1983ia}
B.~Andersson, G.~Gustafson, G.~Ingelman, and T.~Sjostrand, ``{Parton
  Fragmentation and String Dynamics}'',
  \href{https://doi.org/10.1016/0370-1573(83)90080-7}{{\em Phys. Rept.}
  {\bfseries 97} (1983) 31--145}.

\bibitem{Bzdak:2012an}
A.~Bzdak, V.~Koch, and V.~Skokov, ``{Baryon number conservation and the
  cumulants of the net proton distribution}'',
  \href{https://doi.org/10.1103/PhysRevC.87.014901}{{\em Phys. Rev. C}
  {\bfseries 87} (2013) 014901},
  \href{https://arxiv.org/abs/1203.4529}{{\ttfamily arXiv:1203.4529 [hep-ph]}}.

\bibitem{Braun-Munzinger:2020jbk}
P.~Braun-Munzinger, B.~Friman, K.~Redlich, A.~Rustamov, and J.~Stachel,
  ``{Relativistic nuclear collisions: Establishing a non-critical baseline for
  fluctuation measurements}'',
  \href{https://doi.org/10.1016/j.nuclphysa.2021.122141}{{\em Nucl. Phys. A}
  {\bfseries 1008} (2021) 122141},
  \href{https://arxiv.org/abs/2007.02463}{{\ttfamily arXiv:2007.02463
  [nucl-th]}}.

\bibitem{Vovchenko:2019kes}
V.~Vovchenko, B.~D\"onigus, and H.~Stoecker, ``{Canonical statistical model
  analysis of p-p , p--Pb, and Pb--Pb collisions at energies available at the
  CERN Large Hadron Collider}'',
  \href{https://doi.org/10.1103/PhysRevC.100.054906}{{\em Phys. Rev. C}
  {\bfseries 100} (2019) 054906},
  \href{https://arxiv.org/abs/1906.03145}{{\ttfamily arXiv:1906.03145
  [hep-ph]}}.

\bibitem{Cleymans:2020fsc}
J.~Cleymans, P.~M. Lo, K.~Redlich, and N.~Sharma, ``{Multiplicity dependence of
  (multi)strange baryons in the canonical ensemble with phase shift
  corrections}'', \href{https://doi.org/10.1103/PhysRevC.103.014904}{{\em Phys.
  Rev. C} {\bfseries 103} (2021) 014904},
  \href{https://arxiv.org/abs/2009.04844}{{\ttfamily arXiv:2009.04844
  [hep-ph]}}.

\bibitem{Letessier:1993hi}
J.~Letessier, A.~Tounsi, U.~W. Heinz, J.~Sollfrank, and J.~Rafelski,
  ``{Strangeness conservation in hot nuclear fireballs}'',
  \href{https://doi.org/10.1103/PhysRevD.51.3408}{{\em Phys. Rev. D} {\bfseries
  51} (1995) 3408--3435},
  \href{https://arxiv.org/abs/hep-ph/9212210}{{\ttfamily
  arXiv:hep-ph/9212210}}.

\bibitem{Bierlich:2022pfr}
C.~Bierlich {\em et~al.}, ``{A comprehensive guide to the physics and usage of
  PYTHIA 8.3}'', \href{https://arxiv.org/abs/2203.11601}{{\ttfamily
  arXiv:2203.11601 [hep-ph]}}.

\bibitem{Skands:2014pea}
P.~Skands, S.~Carrazza, and J.~Rojo, ``{Tuning PYTHIA 8.1: the Monash 2013
  Tune}'', \href{https://doi.org/10.1140/epjc/s10052-014-3024-y}{{\em Eur.
  Phys. J. C} {\bfseries 74} (2014) 3024},
  \href{https://arxiv.org/abs/1404.5630}{{\ttfamily arXiv:1404.5630 [hep-ph]}}.

\bibitem{Christiansen:2015yqa}
J.~Christiansen and P.~Skands, ``{String Formation Beyond Leading Colour}'',
  \href{https://doi.org/10.1007/JHEP08(2015)003}{{\em JHEP} {\bfseries 08}
  (2015) 003}, \href{https://arxiv.org/abs/1505.01681}{{\ttfamily
  arXiv:1505.01681 [hep-ph]}}.

\bibitem{Bierlich:2014xba}
C.~Bierlich, G.~Gustafson, L.~L\"onnblad, and A.~Tarasov, ``{Effects of
  Overlapping Strings in pp Collisions}'',
  \href{https://doi.org/10.1007/JHEP03(2015)148}{{\em JHEP} {\bfseries 03}
  (2015) 148}, \href{https://arxiv.org/abs/1412.6259}{{\ttfamily
  arXiv:1412.6259 [hep-ph]}}.

\bibitem{Nayak:2018xip}
R.~Nayak, S.~Pal, and S.~Dash, ``{Effect of rope hadronization on strangeness
  enhancement in p$-$p collisions at LHC energies}'',
  \href{https://doi.org/10.1103/PhysRevD.100.074023}{{\em Phys. Rev. D}
  {\bfseries 100} (2019) 074023},
  \href{https://arxiv.org/abs/1812.07718}{{\ttfamily arXiv:1812.07718
  [hep-ph]}}.

\bibitem{ALICE:2019nbs}
{\bfseries ALICE} Collaboration, S.~Acharya {\em et~al.}, ``{Global baryon
  number conservation encoded in net-proton fluctuations measured in
  Pb\textendash{}Pb collisions at $\sqrt{s_{\mathrm{NN}}}=2.76$ TeV}'',
  \href{https://doi.org/10.1016/j.physletb.2020.135564}{{\em Phys. Lett. B}
  {\bfseries 807} (2020) 135564},
  \href{https://arxiv.org/abs/1910.14396}{{\ttfamily arXiv:1910.14396
  [nucl-ex]}}.

\bibitem{Braun-Munzinger:2016yjz}
P.~Braun-Munzinger, A.~Rustamov, and J.~Stachel, ``{Bridging the gap between
  event-by-event fluctuation measurements and theory predictions in
  relativistic nuclear collisions}'',
  \href{https://doi.org/10.1016/j.nuclphysa.2017.01.011}{{\em Nucl. Phys. A}
  {\bfseries 960} (2017) 114--130},
  \href{https://arxiv.org/abs/1612.00702}{{\ttfamily arXiv:1612.00702
  [nucl-th]}}.

\bibitem{ALICE:2023ulv}
{\bfseries ALICE} Collaboration, S.~Acharya {\em et~al.}, ``{Measurements of
  chemical potentials in Pb--Pb collisions at $\sqrt{s_{\rm NN}} = 5.02$
  TeV}'', \href{https://doi.org/10.1103/PhysRevLett.133.092301}{{\em Phys. Rev.
  Lett.} {\bfseries 133} (2024) 092301},
  \href{https://arxiv.org/abs/2311.13332}{{\ttfamily arXiv:2311.13332
  [nucl-ex]}}.

\bibitem{ALICE:2008ngc}
{\bfseries ALICE} Collaboration, K.~Aamodt {\em et~al.}, ``{The ALICE
  experiment at the CERN LHC}'',
  \href{https://doi.org/10.1088/1748-0221/3/08/S08002}{{\em JINST} {\bfseries
  3} (2008) S08002}.

\bibitem{ALICE:2014sbx}
{\bfseries ALICE} Collaboration, B.~Abelev {\em et~al.}, ``{Performance of the
  ALICE Experiment at the CERN LHC}'',
  \href{https://doi.org/10.1142/S0217751X14300440}{{\em Int. J. Mod. Phys. A}
  {\bfseries 29} (2014) 1430044},
  \href{https://arxiv.org/abs/1402.4476}{{\ttfamily arXiv:1402.4476
  [nucl-ex]}}.

\bibitem{ALICE:2013axi}
{\bfseries ALICE} Collaboration, E.~Abbas {\em et~al.}, ``{Performance of the
  ALICE VZERO system}'',
  \href{https://doi.org/10.1088/1748-0221/8/10/P10016}{{\em JINST} {\bfseries
  8} (2013) P10016}, \href{https://arxiv.org/abs/1306.3130}{{\ttfamily
  arXiv:1306.3130 [nucl-ex]}}.

\bibitem{Arslandok:2022dyb}
M.~Arslandok, E.~Hellb\"ar, M.~Ivanov, R.~H. M\"unzer, and J.~Wiechula,
  ``{Track Reconstruction in a High-Density Environment with ALICE}'',
  \href{https://doi.org/10.3390/particles5010008}{{\em Particles} {\bfseries 5}
  (2022) 84--95}, \href{https://arxiv.org/abs/2203.10325}{{\ttfamily
  arXiv:2203.10325 [physics.ins-det]}}.

\bibitem{ALICE:2018tvk}
{\bfseries ALICE} Collaboration, ``{Centrality determination in heavy ion
  collisions}'', {\em ALICE-PUBLIC-2018-011} (2018) .
  \url{https://cds.cern.ch/record/2636623}.

\bibitem{ALICE:2010tia}
{\bfseries ALICE} Collaboration, K.~Aamodt {\em et~al.}, ``{Alignment of the
  ALICE Inner Tracking System with cosmic-ray tracks}'',
  \href{https://doi.org/10.1088/1748-0221/5/03/P03003}{{\em JINST} {\bfseries
  5} (2010) P03003}, \href{https://arxiv.org/abs/1001.0502}{{\ttfamily
  arXiv:1001.0502 [physics.ins-det]}}.

\bibitem{Alme:2010ke}
J.~Alme {\em et~al.}, ``{The ALICE TPC, a large 3-dimensional tracking device
  with fast readout for ultra-high multiplicity events}'',
  \href{https://doi.org/10.1016/j.nima.2010.04.042}{{\em Nucl. Instrum. Meth.
  A} {\bfseries 622} (2010) 316--367},
  \href{https://arxiv.org/abs/1001.1950}{{\ttfamily arXiv:1001.1950
  [physics.ins-det]}}.

\bibitem{ALICE:2013wgn}
{\bfseries ALICE} Collaboration, B.~Abelev {\em et~al.}, ``{Multiplicity
  Dependence of Pion, Kaon, Proton and Lambda Production in p--Pb Collisions at
  $\sqrt{s_\mathrm{NN}}$ = 5.02 TeV}'',
  \href{https://doi.org/10.1016/j.physletb.2013.11.020}{{\em Phys. Lett. B}
  {\bfseries 728} (2014) 25--38},
  \href{https://arxiv.org/abs/1307.6796}{{\ttfamily arXiv:1307.6796
  [nucl-ex]}}.

\bibitem{ALICE:2019hno}
{\bfseries ALICE} Collaboration, S.~Acharya {\em et~al.}, ``{Production of
  charged pions, kaons, and (anti-)protons in Pb--Pb and inelastic pp
  collisions at $\sqrt {s_\mathrm{NN}}$ = 5.02 TeV}'',
  \href{https://doi.org/10.1103/PhysRevC.101.044907}{{\em Phys. Rev. C}
  {\bfseries 101} (2020) 044907},
  \href{https://arxiv.org/abs/1910.07678}{{\ttfamily arXiv:1910.07678
  [nucl-ex]}}.

\bibitem{ALICE:2012yqk}
{\bfseries ALICE} Collaboration, B.~Abelev {\em et~al.}, ``{Multi-strange
  baryon production in pp collisions at $\sqrt{s} = 7$ TeV with ALICE}'',
  \href{https://doi.org/10.1016/j.physletb.2012.05.011}{{\em Phys. Lett. B}
  {\bfseries 712} (2012) 309--318},
  \href{https://arxiv.org/abs/1204.0282}{{\ttfamily arXiv:1204.0282
  [nucl-ex]}}.

\bibitem{ALICE:2013xmt}
{\bfseries ALICE} Collaboration, B.~Abelev {\em et~al.}, ``{Multi-strange
  baryon production at mid-rapidity in Pb--Pb collisions at
  $\sqrt{s_{\rm{NN}}}$ = 2.76 TeV}'',
  \href{https://doi.org/10.1016/j.physletb.2014.05.052}{{\em Phys. Lett. B}
  {\bfseries 728} (2014) 216--227},
  \href{https://arxiv.org/abs/1307.5543}{{\ttfamily arXiv:1307.5543
  [nucl-ex]}}. [Erratum: Phys.Lett.B 734, 409--410 (2014)].

\bibitem{ALICE:2015mpp}
{\bfseries ALICE} Collaboration, J.~Adam {\em et~al.}, ``{Multi-strange baryon
  production in p--Pb collisions at $\sqrt{s_\mathrm{NN}}=5.02$ TeV}'',
  \href{https://doi.org/10.1016/j.physletb.2016.05.027}{{\em Phys. Lett. B}
  {\bfseries 758} (2016) 389--401},
  \href{https://arxiv.org/abs/1512.07227}{{\ttfamily arXiv:1512.07227
  [nucl-ex]}}.

\bibitem{ALICE-PUBLIC-2023-003}
{\bfseries ALICE} Collaboration, ``{Chasing the onset of strange-quark
  thermalization at the LHC}'', {\em ALICE-PUBLIC-2023-003} (2023) .
  \url{https://cds.cern.ch/record/2869064}.

\bibitem{Nonaka:2017kko}
T.~Nonaka, M.~Kitazawa, and S.~Esumi, ``{More efficient formulas for efficiency
  correction of cumulants and effect of using averaged efficiency}'',
  \href{https://doi.org/10.1103/PhysRevC.95.064912}{{\em Phys. Rev. C}
  {\bfseries 95} (2017) 064912},
  \href{https://arxiv.org/abs/1702.07106}{{\ttfamily arXiv:1702.07106
  [physics.data-an]}}. [Erratum: Phys.Rev.C 103, 029901 (2021)].

\bibitem{Pierog:2013ria}
T.~Pierog, I.~Karpenko, J.~M. Katzy, E.~Yatsenko, and K.~Werner, ``{EPOS LHC:
  Test of collective hadronization with data measured at the CERN Large Hadron
  Collider}'', \href{https://doi.org/10.1103/PhysRevC.92.034906}{{\em Phys.
  Rev. C} {\bfseries 92} (2015) 034906},
  \href{https://arxiv.org/abs/1306.0121}{{\ttfamily arXiv:1306.0121 [hep-ph]}}.

\bibitem{Wang:1991hta}
X.-N. Wang and M.~Gyulassy, ``{HIJING: A Monte Carlo model for multiple jet
  production in pp, pA and AA collisions}'',
  \href{https://doi.org/10.1103/PhysRevD.44.3501}{{\em Phys. Rev. D} {\bfseries
  44} (1991) 3501--3516}.

\bibitem{GEANT4:2002zbu}
{\bfseries $\textsc{geant4}$} Collaboration, S.~Agostinelli {\em et~al.},
  ``{$\textsc{geant4}$--a simulation toolkit}'',
  \href{https://doi.org/10.1016/S0168-9002(03)01368-8}{{\em Nucl. Instrum.
  Meth. A} {\bfseries 506} (2003) 250--303}.

\bibitem{Pandav:2018bdx}
A.~Pandav, D.~Mallick, and B.~Mohanty, ``{Effect of limited statistics on
  higher order cumulants measurement in heavy-ion collision experiments}'',
  \href{https://doi.org/10.1016/j.nuclphysa.2019.08.002}{{\em Nucl. Phys. A}
  {\bfseries 991} (2019) 121608},
  \href{https://arxiv.org/abs/1809.08892}{{\ttfamily arXiv:1809.08892
  [nucl-ex]}}.

\bibitem{ALICE:2015juo}
{\bfseries ALICE} Collaboration, J.~Adam {\em et~al.}, ``{Centrality Dependence
  of the Charged-Particle Multiplicity Density at Midrapidity in Pb-Pb
  Collisions at $\sqrt{s_{\rm NN}}$ = 5.02 TeV}'',
  \href{https://doi.org/10.1103/PhysRevLett.116.222302}{{\em Phys. Rev. Lett.}
  {\bfseries 116} (2016) 222302},
  \href{https://arxiv.org/abs/1512.06104}{{\ttfamily arXiv:1512.06104
  [nucl-ex]}}.

\bibitem{ALICE:2021ovi}
{\bfseries ALICE} Collaboration, S.~Acharya {\em et~al.}, ``{Production of
  light (anti)nuclei in pp collisions at $\sqrt{s} = 5.02$~TeV}'',
  \href{https://doi.org/10.1140/epjc/s10052-022-10241-z}{{\em Eur. Phys. J. C}
  {\bfseries 82} (2022) 289},
  \href{https://arxiv.org/abs/2112.00610}{{\ttfamily arXiv:2112.00610
  [nucl-ex]}}.

\bibitem{Vovchenko:2019pjl}
V.~Vovchenko and H.~Stoecker, ``{Thermal-FIST: A package for heavy-ion
  collisions and hadronic equation of state}'',
  \href{https://doi.org/10.1016/j.cpc.2019.06.024}{{\em Comput. Phys. Commun.}
  {\bfseries 244} (2019) 295--310},
  \href{https://arxiv.org/abs/1901.05249}{{\ttfamily arXiv:1901.05249
  [nucl-th]}}.

\bibitem{Bierlich:2018xfw}
C.~Bierlich, G.~Gustafson, L.~L\"onnblad, and H.~Shah, ``{The Angantyr model
  for Heavy-Ion Collisions in PYTHIA8}'',
  \href{https://doi.org/10.1007/JHEP10(2018)134}{{\em JHEP} {\bfseries 10}
  (2018) 134}, \href{https://arxiv.org/abs/1806.10820}{{\ttfamily
  arXiv:1806.10820 [hep-ph]}}.

\bibitem{Vovchenko:2020kwg}
V.~Vovchenko and V.~Koch, ``{Particlization of an interacting hadron resonance
  gas with global conservation laws for event-by-event fluctuations in
  heavy-ion collisions}'',
  \href{https://doi.org/10.1103/PhysRevC.103.044903}{{\em Phys. Rev. C}
  {\bfseries 103} (2021) 044903},
  \href{https://arxiv.org/abs/2012.09954}{{\ttfamily arXiv:2012.09954
  [hep-ph]}}.

\bibitem{ALICE:2022xiu}
{\bfseries ALICE} Collaboration, S.~Acharya {\em et~al.}, ``{First Measurement
  of Antideuteron Number Fluctuations at Energies Available at the Large Hadron
  Collider}'', \href{https://doi.org/10.1103/PhysRevLett.131.041901}{{\em Phys.
  Rev. Lett.} {\bfseries 131} (2023) 041901},
  \href{https://arxiv.org/abs/2204.10166}{{\ttfamily arXiv:2204.10166
  [nucl-ex]}}.

\bibitem{Sun:2022rjh}
K.-J. Sun and C.~M. Ko, ``{Event-by-event antideuteron multiplicity fluctuation
  in Pb+Pb collisions at $\sqrt{s_\mathrm{NN}}=5.02$ TeV}'',
  \href{https://doi.org/10.1016/j.physletb.2023.137864}{{\em Phys. Lett. B}
  {\bfseries 840} (2023) 137864},
  \href{https://arxiv.org/abs/2204.10879}{{\ttfamily arXiv:2204.10879
  [nucl-th]}}.

\bibitem{Gupta:2022phu}
S.~Gupta, D.~Mallick, D.~K. Mishra, B.~Mohanty, and N.~Xu, ``{Limits of
  thermalization in relativistic heavy ion collisions}'',
  \href{https://doi.org/10.1016/j.physletb.2022.137021}{{\em Phys. Lett. B}
  {\bfseries 829} (2022) 137021}.

\bibitem{Bazavov:2017dus}
A.~Bazavov {\em et~al.}, ``{The QCD Equation of State to $\mathcal{O}(\mu_B^6)$
  from Lattice QCD}'', \href{https://doi.org/10.1103/PhysRevD.95.054504}{{\em
  Phys. Rev. D} {\bfseries 95} (2017) 054504},
  \href{https://arxiv.org/abs/1701.04325}{{\ttfamily arXiv:1701.04325
  [hep-lat]}}.

\bibitem{Braun-Munzinger:2023gsd}
P.~Braun-Munzinger, K.~Redlich, A.~Rustamov, and J.~Stachel, ``{The imprint of
  conservation laws on correlated particle production}'',
  \href{https://doi.org/10.1007/JHEP08(2024)113}{{\em JHEP} {\bfseries 08}
  (2024) 113}, \href{https://arxiv.org/abs/2312.15534}{{\ttfamily
  arXiv:2312.15534 [nucl-th]}}.

\bibitem{ALICE:2022xpf}
{\bfseries ALICE} Collaboration, S.~Acharya {\em et~al.}, ``{Closing in on
  critical net-baryon fluctuations at LHC energies: Cumulants up to third order
  in Pb\textendash{}Pb collisions}'',
  \href{https://doi.org/10.1016/j.physletb.2022.137545}{{\em Phys. Lett. B}
  {\bfseries 844} (2023) 137545},
  \href{https://arxiv.org/abs/2206.03343}{{\ttfamily arXiv:2206.03343
  [nucl-ex]}}.

\end{thebibliography}\endgroup


\newpage
\appendix

\section{EndMatter}

\subsection{Systematic uncertainties}\label{app:sys}
The systematic uncertainties on both $\rho_{\Delta\Xi\Delta\mathrm{K}}$ and $\kappa_2(\Delta\Xi)/\kappa_1(\overline{\Xi}^+ + \Xi^-)$ are summarised in \cref{tab:sys}. The contributions associated to the different sources are separately reported. The sources considered for charged kaon candidates are the selection on the number of TPC space points per track, the TPC and TOF PID criteria, the selection on the distance of closest approach (DCA) to the primary vertex (PV) of the reconstructed tracks, and the track $\chi^2$ selection for charged kaons. For the $\Xi$ candidates, the systematic sources are the applied BDT threshold and the invariant-mass selection.

\begin{table}[!htbp]
\begin{center}
\caption{Relative systematic uncertainties on the event-by-event observables due to the different sources considered in this Letter. Only the contributions relevant to the employed particle species are reported.}
\begin{tabular}{ l c c } \hline
Source & $\rho_{\Delta\Xi\Delta\mathrm{K}}$ & $\kappa_2(\Delta\Xi)/\kappa_1(\overline{\Xi}^+ + \Xi^-)$ \\ \hline
TPC space points & 0.6\% & -- \\
PID selections& 0.6\% & -- \\
DCA to PV & 0.6\% & -- \\
Track $\chi^2$ & 0.3\% & -- \\
$\Xi$ mass selection & 1\% & 0.3\% \\
BDT & 1\% & 0.2\% \\ \hline
\end{tabular}
\label{tab:sys}
\end{center}
\end{table}

\subsection{V$_c$ determination}
\label{app:vc_determination}
\begin{figure}[h]
    \centering
    \includegraphics[width=0.49\textwidth]{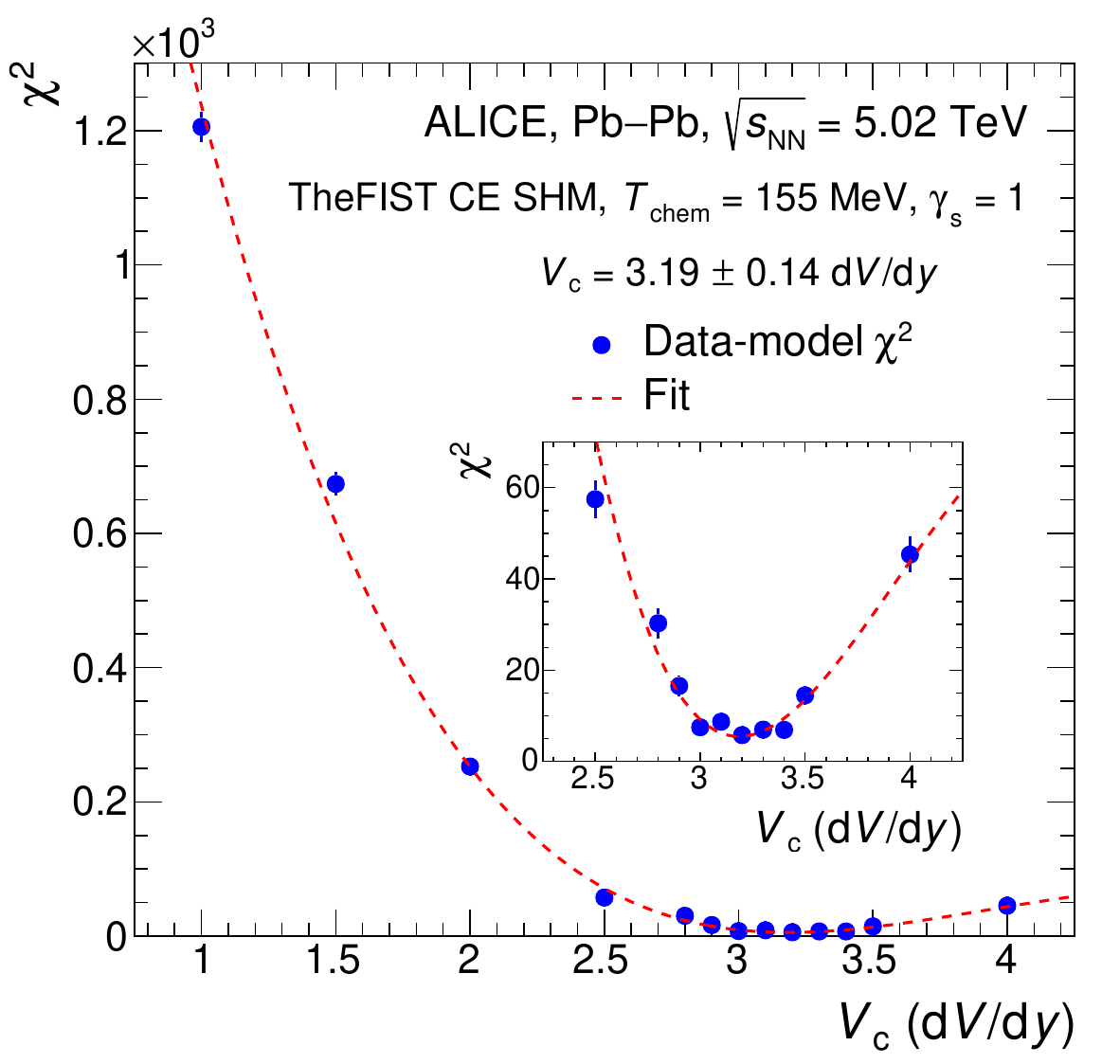}
    \caption{
    Combined $\chi^2$ profile for the extraction of the correlation volume for baryon and strangeness conservation obtained by comparing the normalized second order net-$\Xi$ cumulants and the $\rho_{\Delta\Xi\Delta K}$ in Pb--Pb collisions to the Thermal-FIST predictions. The region close to the minimum $\chi^2$ is shown in the inset. The dashed line represents a fit to the $\chi^2$ profile with a fourth-degree polynomial function.}
    \label{fig:chi2_profile}
\end{figure}

The best estimate of the correlation volume, $V_{\mathrm{c}}$, for quantum number conservation in CE SHM is extracted from the data by comparing the experimental results to SHM model predictions based on different values of $V_{\mathrm{c}}$. This study is performed in the multiplicity region corresponding to Pb--Pb collisions, where the approximation $\gamma_s\approx 1$ holds, by varying the correlation volume in the range $1.0\leq V_{\mathrm{c}}\leq4.0\ \mathrm{d}V/\mathrm{d}y$ with a step of $\Delta V_{\mathrm{c}}=0.5\ \mathrm{d}V/\mathrm{d}y$~\cite{ALICE-PUBLIC-2023-003}. The step is decreased close to the minimum of the $\chi^2$ profile to obtain a better sampling in that region. The chemical freeze-out temperature is set to $T_\mathrm{chem}=155\ \mathrm{MeV}$. The quantum numbers that are conserved over $V_{\mathrm{c}}$ are the baryon number, $B$, and strangeness, $S$. The latter has a larger effect on the magnitude of the anticorrelation between the net-kaon and net-$\Xi$ numbers. The agreement between the experimental measurements and the model predictions is quantified by a combined $\chi^{2}$ that simultaneously accounts for the discrepancy between model predictions and data for both the analyzed observables. The 6 multiplicity intervals corresponding to the results in Pb--Pb collisions are used in this comparison, consisting of 12 data points in total for $\rho_{\Delta \Xi\Delta\mathrm{K}}$ and the $\kappa_2/\kappa_1$ ratio. The $\chi^2$ calculation takes into account only the statistical uncertainties. The obtained $\chi^2$ profile as a function of $V_{\mathrm{c}}$ is shown in \cref{fig:chi2_profile}. The observed $\chi^{2}$ is dominated by $\rho_{\Delta \Xi\Delta\mathrm{K}}$ because it has smaller uncertainties with respect to the $\kappa_2/\kappa_1$ ratio of the net-$\Xi$ number. The systematic uncertainties on $\rho_{\Delta \Xi\Delta\mathrm{K}}$ and $\kappa_2/\kappa_1(\Delta\Xi)$ are also included in the $V_{\mathrm{c}}$ evaluation assuming full correlation of the systematic uncertainties with multiplicity. The assigned uncertainty is obtained as half of the difference between the $V_{\mathrm{c}}$ values obtained by repeating the fit, shifting upward and downward the event-by-event observables by their systematic uncertainties. The final uncertainty on $V_{\mathrm{c}}$ is obtained by adding the statistical and systematic contributions in quadrature. The correlation volume obtained from the minimization of the observed $\chi^2$ profile is $V_{\mathrm{c}}=3.19\pm0.14\ \mathrm{d}V/\mathrm{d}y$, with a fit probability of $P=0.94$.

\subsection{Comparison to net-proton fluctuations}
\begin{figure}[h]
    \centering
    \includegraphics[width=0.49\textwidth]{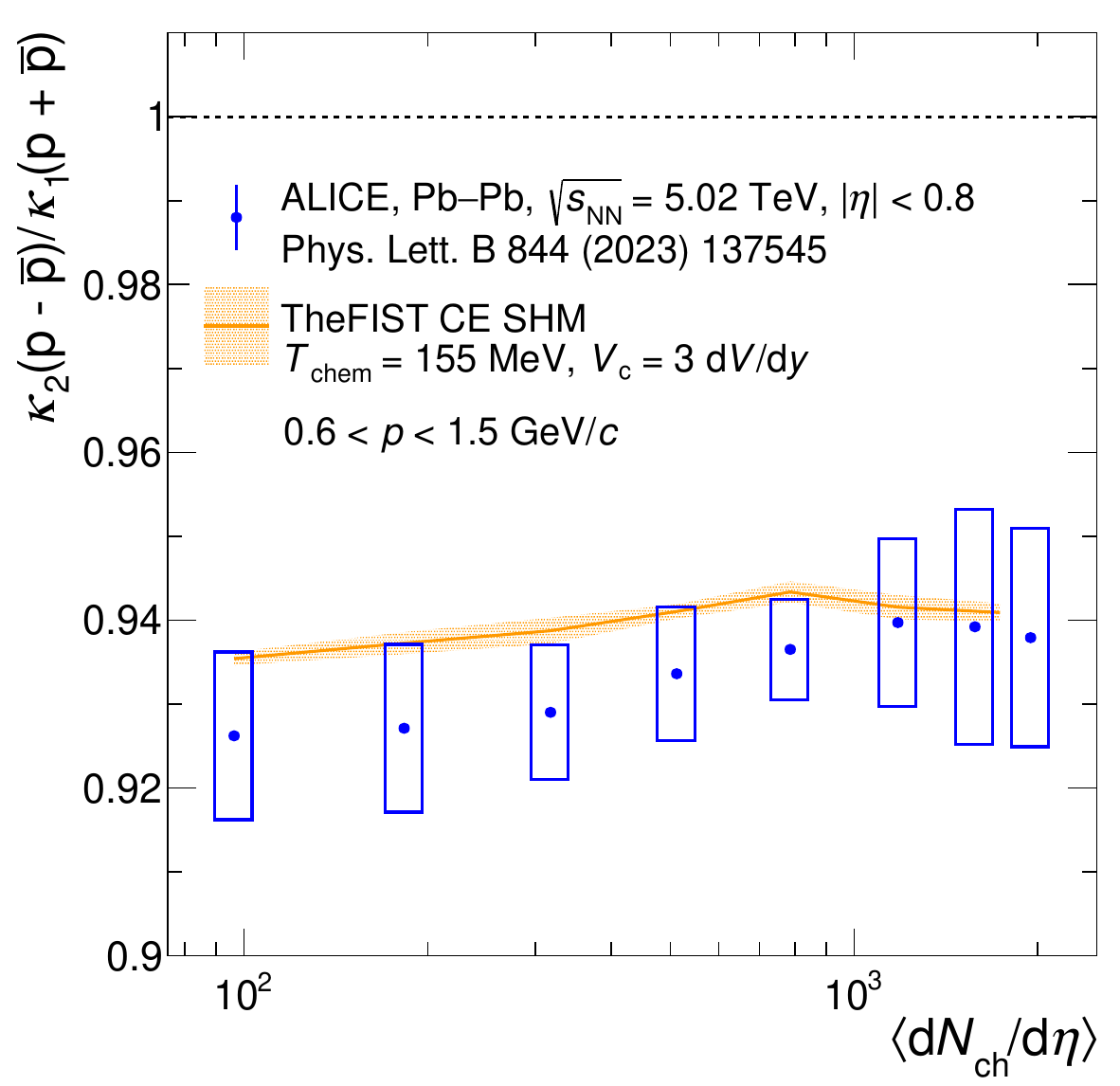}
    \caption{
    Comparison between the net-proton normalized second-order cumulant as a function of the average multiplicity at midrapidity measured by the ALICE Collaboration in Pb--Pb collisions at $\sqrt{s_{\mathrm{NN}}}=5.02$~TeV (blue circles)~\cite{ALICE:2022xpf} and the predictions of the Thermal-FIST model, with $T_{\mathrm{chem}}=155$~MeV and $V_{\mathrm{c}}=3\ \mathrm{d}V/\mathrm{dy}$ (orange band).}
    \label{fig:netproton}
\end{figure}

In~\cref{fig:netproton}, the comparison between the published $\kappa_2/\kappa_1$ normalized second-order cumulant of net-protons and the model calculations obtained with the Thermal-FIST model is shown. In the model parametrization, the chemical freeze-out temperature and the strangeness saturation factor are set to $T_{\mathrm{chem}}=155\ \mathrm{MeV}$ and $\gamma_{s}=1$, respectively. The parametrization of the system volume per unit of rapidity, $\mathrm{d}V/\mathrm{d}y$, is obtained from~Ref.~\cite{Vovchenko:2019kes}.

\label{app:netproton}
%
\newpage
\section{The ALICE Collaboration}
\label{app:collab}
\begin{flushleft} 
\small

S.~Acharya\,\orcidlink{0000-0002-9213-5329}\,$^{\rm 127}$, 
D.~Adamov\'{a}\,\orcidlink{0000-0002-0504-7428}\,$^{\rm 86}$, 
A.~Agarwal$^{\rm 135}$, 
G.~Aglieri Rinella\,\orcidlink{0000-0002-9611-3696}\,$^{\rm 32}$, 
L.~Aglietta\,\orcidlink{0009-0003-0763-6802}\,$^{\rm 24}$, 
M.~Agnello\,\orcidlink{0000-0002-0760-5075}\,$^{\rm 29}$, 
N.~Agrawal\,\orcidlink{0000-0003-0348-9836}\,$^{\rm 25}$, 
Z.~Ahammed\,\orcidlink{0000-0001-5241-7412}\,$^{\rm 135}$, 
S.~Ahmad\,\orcidlink{0000-0003-0497-5705}\,$^{\rm 15}$, 
S.U.~Ahn\,\orcidlink{0000-0001-8847-489X}\,$^{\rm 71}$
% ...

\section*{Affiliation Notes}

$^{\rm I}$ Deceased\\
$^{\rm II}$ Also at: Max-Planck-Institut fur Physik, Munich, Germany\\
$^{\rm III}$ Also at: Italian National Agency for New Technologies, Energy and Sustainable Economic Development (ENEA), Bologna, Italy\\
$^{\rm IV}$ Also at: Dipartimento DET del Politecnico di Torino, Turin, Italy\\
$^{\rm V}$ Also at: Yildiz Technical University, Istanbul, T\"{u}rkiye\\
$^{\rm VI}$ Also at: Department of Applied Physics, Aligarh Muslim University, Aligarh, India\\
$^{\rm VII}$ Also at: Institute of Theoretical Physics, University of Wroclaw, Poland\\
$^{\rm VIII}$ Also at: An institution covered by a cooperation agreement with CERN\\

\section*{Collaboration Institutes}

$^{1}$ A.I. Alikhanyan National Science Laboratory (Yerevan Physics Institute) Foundation, Yerevan, Armenia\\
$^{2}$ AGH University of Krakow, Cracow, Poland\\
$^{3}$ Bogolyubov Institute for Theoretical Physics, National Academy of Sciences of Ukraine, Kiev, Ukraine\\
$^{4}$ Bose Institute, Department of Physics  and Centre for Astroparticle Physics and Space Science (CAPSS), Kolkata, India\\
$^{5}$ California Polytechnic State University, San Luis Obispo, California, United States\\
$^{6}$ Central China Normal University, Wuhan, China\\
$^{7}$ Centro de Aplicaciones Tecnol\'{o}gicas y Desarrollo Nuclear (CEADEN), Havana, Cuba\\
$^{8}$ Centro de Investigaci\'{o}n y de Estudios Avanzados (CINVESTAV), Mexico City and M\'{e}rida, Mexico\\
$^{9}$ Chicago State University, Chicago, Illinois, United States\\
$^{10}$ China Institute of Atomic Energy, Beijing, China\\
$^{11}$ China University of Geosciences, Wuhan, China\\
$^{12}$ Chungbuk National University, Cheongju, Republic of Korea\\
$^{13}$ Comenius University Bratislava, Faculty of Mathematics, Physics and Informatics, Bratislava, Slovak Republic\\
$^{14}$ Creighton University, Omaha, Nebraska, United States\\
$^{15}$ Department of Physics, Aligarh Muslim University, Aligarh, India\\
$^{16}$ Department of Physics, Pusan National University, Pusan, Republic of Korea\\
$^{17}$ Department of Physics, Sejong University, Seoul, Republic of Korea\\
$^{18}$ Department of Physics, University of California, Berkeley, California, United States\\
$^{19}$ Department of Physics, University of Oslo, Oslo, Norway\\
$^{20}$ Department of Physics and Technology, University of Bergen, Bergen, Norway\\
$^{21}$ Dipartimento di Fisica, Universit\`{a} di Pavia, Pavia, Italy\\
$^{22}$ Dipartimento di Fisica dell'Universit\`{a} and Sezione INFN, Cagliari, Italy\\
$^{23}$ Dipartimento di Fisica dell'Universit\`{a} and Sezione INFN, Trieste, Italy\\
$^{24}$ Dipartimento di Fisica dell'Universit\`{a} and Sezione INFN, Turin, Italy\\
$^{25}$ Dipartimento di Fisica e Astronomia dell'Universit\`{a} and Sezione INFN, Bologna, Italy\\
$^{26}$ Dipartimento di Fisica e Astronomia dell'Universit\`{a} and Sezione INFN, Catania, Italy\\
$^{27}$ Dipartimento di Fisica e Astronomia dell'Universit\`{a} and Sezione INFN, Padova, Italy\\
$^{28}$ Dipartimento di Fisica `E.R.~Caianiello' dell'Universit\`{a} and Gruppo Collegato INFN, Salerno, Italy\\
$^{29}$ Dipartimento DISAT del Politecnico and Sezione INFN, Turin, Italy\\
$^{30}$ Dipartimento di Scienze MIFT, Universit\`{a} di Messina, Messina, Italy\\
$^{31}$ Dipartimento Interateneo di Fisica `M.~Merlin' and Sezione INFN, Bari, Italy\\
$^{32}$ European Organization for Nuclear Research (CERN), Geneva, Switzerland\\
$^{33}$ Faculty of Electrical Engineering, Mechanical Engineering and Naval Architecture, University of Split, Split, Croatia\\
$^{34}$ Faculty of Engineering and Science, Western Norway University of Applied Sciences, Bergen, Norway\\
$^{35}$ Faculty of Nuclear Sciences and Physical Engineering, Czech Technical University in Prague, Prague, Czech Republic\\
$^{36}$ Faculty of Physics, Sofia University, Sofia, Bulgaria\\
$^{37}$ Faculty of Science, P.J.~\v{S}af\'{a}rik University, Ko\v{s}ice, Slovak Republic\\
$^{38}$ Frankfurt Institute for Advanced Studies, Johann Wolfgang Goethe-Universit\"{a}t Frankfurt, Frankfurt, Germany\\
$^{39}$ Fudan University, Shanghai, China\\
$^{40}$ Gangneung-Wonju National University, Gangneung, Republic of Korea\\
$^{41}$ Gauhati University, Department of Physics, Guwahati, India\\
$^{42}$ Helmholtz-Institut f\"{u}r Strahlen- und Kernphysik, Rheinische Friedrich-Wilhelms-Universit\"{a}t Bonn, Bonn, Germany\\
$^{43}$ Helsinki Institute of Physics (HIP), Helsinki, Finland\\
$^{44}$ High Energy Physics Group,  Universidad Aut\'{o}noma de Puebla, Puebla, Mexico\\
$^{45}$ Horia Hulubei National Institute of Physics and Nuclear Engineering, Bucharest, Romania\\
$^{46}$ HUN-REN Wigner Research Centre for Physics, Budapest, Hungary\\
$^{47}$ Indian Institute of Technology Bombay (IIT), Mumbai, India\\
$^{48}$ Indian Institute of Technology Indore, Indore, India\\
$^{49}$ INFN, Laboratori Nazionali di Frascati, Frascati, Italy\\
$^{50}$ INFN, Sezione di Bari, Bari, Italy\\
$^{51}$ INFN, Sezione di Bologna, Bologna, Italy\\
$^{52}$ INFN, Sezione di Cagliari, Cagliari, Italy\\
$^{53}$ INFN, Sezione di Catania, Catania, Italy\\
$^{54}$ INFN, Sezione di Padova, Padova, Italy\\
$^{55}$ INFN, Sezione di Pavia, Pavia, Italy\\
$^{56}$ INFN, Sezione di Torino, Turin, Italy\\
$^{57}$ INFN, Sezione di Trieste, Trieste, Italy\\
$^{58}$ Inha University, Incheon, Republic of Korea\\
$^{59}$ Institute for Gravitational and Subatomic Physics (GRASP), Utrecht University/Nikhef, Utrecht, Netherlands\\
$^{60}$ Institute of Experimental Physics, Slovak Academy of Sciences, Ko\v{s}ice, Slovak Republic\\
$^{61}$ Institute of Physics, Homi Bhabha National Institute, Bhubaneswar, India\\
$^{62}$ Institute of Physics of the Czech Academy of Sciences, Prague, Czech Republic\\
$^{63}$ Institute of Space Science (ISS), Bucharest, Romania\\
$^{64}$ Institut f\"{u}r Kernphysik, Johann Wolfgang Goethe-Universit\"{a}t Frankfurt, Frankfurt, Germany\\
$^{65}$ Instituto de Ciencias Nucleares, Universidad Nacional Aut\'{o}noma de M\'{e}xico, Mexico City, Mexico\\
$^{66}$ Instituto de F\'{i}sica, Universidade Federal do Rio Grande do Sul (UFRGS), Porto Alegre, Brazil\\
$^{67}$ Instituto de F\'{\i}sica, Universidad Nacional Aut\'{o}noma de M\'{e}xico, Mexico City, Mexico\\
$^{68}$ iThemba LABS, National Research Foundation, Somerset West, South Africa\\
$^{69}$ Jeonbuk National University, Jeonju, Republic of Korea\\
$^{70}$ Johann-Wolfgang-Goethe Universit\"{a}t Frankfurt Institut f\"{u}r Informatik, Fachbereich Informatik und Mathematik, Frankfurt, Germany\\
$^{71}$ Korea Institute of Science and Technology Information, Daejeon, Republic of Korea\\
$^{72}$ KTO Karatay University, Konya, Turkey\\
$^{73}$ Laboratoire de Physique Subatomique et de Cosmologie, Universit\'{e} Grenoble-Alpes, CNRS-IN2P3, Grenoble, France\\
$^{74}$ Lawrence Berkeley National Laboratory, Berkeley, California, United States\\
$^{75}$ Lund University Department of Physics, Division of Particle Physics, Lund, Sweden\\
$^{76}$ Nagasaki Institute of Applied Science, Nagasaki, Japan\\
$^{77}$ Nara Women{'}s University (NWU), Nara, Japan\\
$^{78}$ National and Kapodistrian University of Athens, School of Science, Department of Physics , Athens, Greece\\
$^{79}$ National Centre for Nuclear Research, Warsaw, Poland\\
$^{80}$ National Institute of Science Education and Research, Homi Bhabha National Institute, Jatni, India\\
$^{81}$ National Nuclear Research Center, Baku, Azerbaijan\\
$^{82}$ National Research and Innovation Agency - BRIN, Jakarta, Indonesia\\
$^{83}$ Niels Bohr Institute, University of Copenhagen, Copenhagen, Denmark\\
$^{84}$ Nikhef, National institute for subatomic physics, Amsterdam, Netherlands\\
$^{85}$ Nuclear Physics Group, STFC Daresbury Laboratory, Daresbury, United Kingdom\\
$^{86}$ Nuclear Physics Institute of the Czech Academy of Sciences, Husinec-\v{R}e\v{z}, Czech Republic\\
$^{87}$ Oak Ridge National Laboratory, Oak Ridge, Tennessee, United States\\
$^{88}$ Ohio State University, Columbus, Ohio, United States\\
$^{89}$ Physics department, Faculty of science, University of Zagreb, Zagreb, Croatia\\
$^{90}$ Physics Department, Panjab University, Chandigarh, India\\
$^{91}$ Physics Department, University of Jammu, Jammu, India\\
$^{92}$ Physics Program and International Institute for Sustainability with Knotted Chiral Meta Matter (SKCM2), Hiroshima University, Hiroshima, Japan\\
$^{93}$ Physikalisches Institut, Eberhard-Karls-Universit\"{a}t T\"{u}bingen, T\"{u}bingen, Germany\\
$^{94}$ Physikalisches Institut, Ruprecht-Karls-Universit\"{a}t Heidelberg, Heidelberg, Germany\\
$^{95}$ Physik Department, Technische Universit\"{a}t M\"{u}nchen, Munich, Germany\\
$^{96}$ Politecnico di Bari and Sezione INFN, Bari, Italy\\
$^{97}$ Research Division and ExtreMe Matter Institute EMMI, GSI Helmholtzzentrum f\"ur Schwerionenforschung GmbH, Darmstadt, Germany\\
$^{98}$ Saga University, Saga, Japan\\
$^{99}$ Saha Institute of Nuclear Physics, Homi Bhabha National Institute, Kolkata, India\\
$^{100}$ School of Physics and Astronomy, University of Birmingham, Birmingham, United Kingdom\\
$^{101}$ Secci\'{o}n F\'{\i}sica, Departamento de Ciencias, Pontificia Universidad Cat\'{o}lica del Per\'{u}, Lima, Peru\\
$^{102}$ Stefan Meyer Institut f\"{u}r Subatomare Physik (SMI), Vienna, Austria\\
$^{103}$ SUBATECH, IMT Atlantique, Nantes Universit\'{e}, CNRS-IN2P3, Nantes, France\\
$^{104}$ Sungkyunkwan University, Suwon City, Republic of Korea\\
$^{105}$ Suranaree University of Technology, Nakhon Ratchasima, Thailand\\
$^{106}$ Technical University of Ko\v{s}ice, Ko\v{s}ice, Slovak Republic\\
$^{107}$ The Henryk Niewodniczanski Institute of Nuclear Physics, Polish Academy of Sciences, Cracow, Poland\\
$^{108}$ The University of Texas at Austin, Austin, Texas, United States\\
$^{109}$ Universidad Aut\'{o}noma de Sinaloa, Culiac\'{a}n, Mexico\\
$^{110}$ Universidade de S\~{a}o Paulo (USP), S\~{a}o Paulo, Brazil\\
$^{111}$ Universidade Estadual de Campinas (UNICAMP), Campinas, Brazil\\
$^{112}$ Universidade Federal do ABC, Santo Andre, Brazil\\
$^{113}$ Universitatea Nationala de Stiinta si Tehnologie Politehnica Bucuresti, Bucharest, Romania\\
$^{114}$ University of Cape Town, Cape Town, South Africa\\
$^{115}$ University of Derby, Derby, United Kingdom\\
$^{116}$ University of Houston, Houston, Texas, United States\\
$^{117}$ University of Jyv\"{a}skyl\"{a}, Jyv\"{a}skyl\"{a}, Finland\\
$^{118}$ University of Kansas, Lawrence, Kansas, United States\\
$^{119}$ University of Liverpool, Liverpool, United Kingdom\\
$^{120}$ University of Science and Technology of China, Hefei, China\\
$^{121}$ University of South-Eastern Norway, Kongsberg, Norway\\
$^{122}$ University of Tennessee, Knoxville, Tennessee, United States\\
$^{123}$ University of the Witwatersrand, Johannesburg, South Africa\\
$^{124}$ University of Tokyo, Tokyo, Japan\\
$^{125}$ University of Tsukuba, Tsukuba, Japan\\
$^{126}$ Universit\"{a}t M\"{u}nster, Institut f\"{u}r Kernphysik, M\"{u}nster, Germany\\
$^{127}$ Universit\'{e} Clermont Auvergne, CNRS/IN2P3, LPC, Clermont-Ferrand, France\\
$^{128}$ Universit\'{e} de Lyon, CNRS/IN2P3, Institut de Physique des 2 Infinis de Lyon, Lyon, France\\
$^{129}$ Universit\'{e} de Strasbourg, CNRS, IPHC UMR 7178, F-67000 Strasbourg, France, Strasbourg, France\\
$^{130}$ Universit\'{e} Paris-Saclay, Centre d'Etudes de Saclay (CEA), IRFU, D\'{e}partment de Physique Nucl\'{e}aire (DPhN), Saclay, France\\
$^{131}$ Universit\'{e}  Paris-Saclay, CNRS/IN2P3, IJCLab, Orsay, France\\
$^{132}$ Universit\`{a} degli Studi di Foggia, Foggia, Italy\\
$^{133}$ Universit\`{a} del Piemonte Orientale, Vercelli, Italy\\
$^{134}$ Universit\`{a} di Brescia, Brescia, Italy\\
$^{135}$ Variable Energy Cyclotron Centre, Homi Bhabha National Institute, Kolkata, India\\
$^{136}$ Warsaw University of Technology, Warsaw, Poland\\
$^{137}$ Wayne State University, Detroit, Michigan, United States\\
$^{138}$ Yale University, New Haven, Connecticut, United States\\
$^{139}$ Yonsei University, Seoul, Republic of Korea\\
$^{140}$  Zentrum  f\"{u}r Technologie und Transfer (ZTT), Worms, Germany\\
$^{141}$ Affiliated with an institute covered by a cooperation agreement with CERN\\
$^{142}$ Affiliated with an international laboratory covered by a cooperation agreement with CERN.\\

\end{flushleft} 

\end{document}